\documentclass[twoside,twocolumn,english,superscriptaddress,nofootinbib,preprintnumbers, aps]{revtex4}

\usepackage{amsmath}
\usepackage{multirow}
\usepackage{booktabs}
\usepackage{multirow}

\usepackage{graphicx}

\usepackage{float}

\def\p{\pi}

\def\s{\sigma}
\def\S{\Sigma}

\def\s{\sigma}
\def\nn{\nonumber}
\def\p{\partial}
\def\ls{\left[}
\def\rs{\right]}
\def\lc{\left\{}
\def\rc{\right\}}

\newcommand{\bi}{\begin{itemize}}
\newcommand{\ei}{\end{itemize}}

\usepackage{color}
\usepackage{amsmath}
\usepackage{graphicx}
\usepackage{amssymb}
\usepackage{esint}
\usepackage[hyperfootnotes=true,unicode=true, 
 bookmarks=true,bookmarksnumbered=false,bookmarksopen=false,
 breaklinks=false,pdfborder={0 0 1},backref=false,colorlinks=true]
 {hyperref}

\makeatletter

\@ifundefined{textcolor}{}
{%
 \definecolor{BLACK}{gray}{0}
 \definecolor{WHITE}{gray}{1}
 \definecolor{RED}{rgb}{1,0,0}
 \definecolor{GREEN}{rgb}{0,1,0}
 \definecolor{BLUE}{rgb}{0,0,1}
 \definecolor{CYAN}{cmyk}{1,0,0,0}
 \definecolor{MAGENTA}{cmyk}{0,1,0,0}
 \definecolor{YELLOW}{cmyk}{0,0,1,0}
 }

\usepackage{hyperref}
\hypersetup{
    colorlinks,%
    citecolor=blue,%
    filecolor=blue,%
    linkcolor=blue,%
    urlcolor=blue
}

\makeatother

\def\s{\sigma}
\def\nn{\nonumber}
\def\p{\partial}
\def\ls{\left[}
\def\rs{\right]}
\def\lc{\left\{}
\def\rc{\right\}}
\newcommand{\be}{\begin{eqnarray}}
\newcommand{\ee}{\end{eqnarray}}

\begin{document}

\title{Complex Linear Effective Theory and Supersymmetry Breaking Vacua}

\author{Fotis Farakos}
\email{fotisf@mail.muni.cz}
\affiliation{ Institute for Theoretical Physics, Masaryk University, \\  611 37 Brno, Czech Republic}

\author{Rikard von Unge}
\email{unge@physics.muni.cz}
\affiliation{ Institute for Theoretical Physics, Masaryk University, \\  611 37 Brno, Czech Republic}


\begin{abstract}

We calculate the low energy effective action of massless and massive complex linear superfields 
coupled to a massive U(1) vector multiplet. 
Our calculations include superspace higher derivative corrections and therefore 
go beyond previous results. 
Among the superspace higher derivatives we find that terms which lead to a deformation of the auxiliary field potential 
and may break supersymmetry 
are also generated. 
We show that the supersymmetry breaking vacua can only be trusted  
if there exists  a hierarchy between the higher order terms. 
A renormalization group analysis shows that  generically a hierarchy is not generated 
by the quantum corrections.

\end{abstract}



\maketitle

\flushbottom

\section{Introduction}

Supersymmetric theories (see for example Ref. \cite{Gates:1983nr,Buchbinder:1998qv}) are candidates 
for describing physics beyond the electroweak scale. 
Despite their many virtues, 
the observed elementary particle spectrum signals that we live in a vacuum where supersymmetry  
has to be broken, thus leading to a  lift of the mass degeneracy 
of the fermionic and bosonic states. 
In a quantum field theory, to understand the true vacuum structure, 
one has to take into account the quantum corrections, 
which in principle will include all the terms allowed by the symmetries.

For chiral models,  the effective K\"ahler potential and superpotential 
are well known \cite{Grisaru:1982df,Grisaru:1996ve,Buchbinder:1994iw,Buchbinder:1994xq}. 
Effective  theories will also  include higher dimension operators which contain superspace higher derivatives. 
The possible effect of these  superspace higher derivatives 
on the vacuum structure 
was initially investigated in Ref. \cite{CFG}  but no conventional supersymmetry breaking was found.
Indeed, as has been shown in Ref. \cite{KLO,FK1,FK2}, 
even though new branches exist, they have to be sustained by non-trivial background fluxes.
The specific higher derivative superspace operators used in Ref. \cite{CFG,KLO,FK1,FK2} 
were shown to be generated by perturbation theory in the work of 
Ref. \cite{Buchbinder:1994iw,Buchbinder:1994xq,Kuzenko:2007cg}, 
but they were also related to  effective actions originating 
from a more fundamental theory \cite{Cecotti:1986gb,Karlhede:1987bg,GonzalezRey:1998kh,Rocek:1997hi,Tseytlin:1999dj}. 
Note that models of this sort have also been used to describe 
supersymmetric skyrmions \cite{BNS,Gates:1995fx,Gates:1996cq,Adam:2013awa}. 
In Ref. \cite{Farakos:2013zsa} linear and complex linear multiplets have also been studied, 
and the resulting theories had a new broken branch admitting non-linear supersymmetry. 
In these theories there are examples where no background flux is needed to sustain the broken branch.

Another important aspect of supersymmetric theories is the various existing supermultiplets.
The most common in use are the chiral multiplet and the gauge multiplet.
The gauge multiplet is somehow unique.
The same is not true for the chiral multiplet; 
there exist other supermultiplets with the same on-shell field content.
These superfields are classically equivalent but with different quantum properties.
In fact the role of the {\it variant} \cite{Gates:1980az,Deo:1985ix} scalar multiplets has been little appreciated.
It is possible that matter fields are not accommodated into a chiral multiplet, 
but rather into a complex linear multiplet for example, 
and the same goes for the supersymmetry breaking sector.

The quantum properties of complex linear supermultiplets have been studied before
\cite{Grisaru:1997hf,Penati:1997pm,Penati:1997fz,TartaglinoMazzucchelli:2004vt} and it was shown that the general quantization procedure
is rather involved. 
If one solves the constraints by introducing prepotentials one is faced with a theory with a
gauge symmetry that needs to be fixed. In the BV formalism this leads to an infinite tower of ghosts. However, as
was shown in Ref. \cite{Penati:1997pm}, 
if the complex linear superfields prepotentials appear only through their field strengths 
({\em i.e.} the superfields themselves) 
the ghosts decouple and one may ignore many of the complications originating in the infinite ghost sector. 
In Ref. \cite{Penati:1997pm}, the effective K\"ahler potential of the sigma model defined in terms of chiral superfields
or its dual defined in terms of complex linear superfields were shown to remain dual at one loop. 
In this work we go beyond the known results about the effective K\"ahler potential 
and we calculate all the one-loop corrections which also include superspace higher derivatives.

Superspace higher derivatives of complex linear multiplets 
have an intriguing effect: 
they deform the auxiliary field potential in such a way that 
the auxiliary field equations of motion have more than one solutions, 
and the new ones lead to vacua where supersymmetry is spontaneously broken. 
Recently the structure of various theories with this possibility were presented \cite{Farakos:2013zsa}. 
The standard property is the auxiliary structure
\be
{\cal L}_{aux} = -F \bar F + \frac{1}{2 f^2} F^2 \bar F^2   , 
\ee
leading to new non-supersymmetric vacua when the auxiliary field $F$ is integrated out.
Indeed on top of the standard solution
\be
F=0 , 
\ee
now exists a second solution to the equations of motion of $F$ 
\be
F \bar F =  f^2 , 
\ee 
which implies supersymmetry breaking. 
Moreover, in a generic study on supersymmetric effective theories one should take into account 
these terms since they  deform the auxiliary field potential and 
thus have an indispensable effect on the scalar potential as well.  
It is then clear that an effective theory should not be restricted to the evaluation 
of the form of the effective K\"ahler potential only, 
but also to the evaluation of the superspace higher derivative operators which are always generated.

\vspace{0.3cm}

Our work is organized as follows: In the next section we review  basic formalism about classical 
and quantum superfields, and present the scalar multiplets variant pictures. 
In section 3 we work with a massless complex linear multiplet coupled to a massive $U(1)$, 
which we integrate out  
and calculate the low energy effective theory and the deformation in the auxiliary field potential. 
In section 4 we work with a massive complex linear multiplet coupled to a massive $U(1)$, 
and by integrating out the massive sector we calculate the low energy effective theory,  
we study the vacuum structure  and 
the flow of the  supersymmetry breaking operators under the renormalization group. 
We conclude with a short discussion  in section 5.

\section{Short review of superfield methods}

Here we review known results in 4-D, ${\cal N}=1$ superspace 
concerning some classical and quantum properties, 
which we will use throughout our work. 
For our conventions one may consult Ref. \cite{Gates:1983nr}.

\subsection{Scalar multiplets and hypermultiplets}

In supersymmetric field theories there exist different ways of 
introducing physical scalar fields into supermultiplets.
These different ways and their equivalence can be understood in terms of dualities between the various superfields.
For example a chiral superfield is defined by the condition
\be
\bar D_{\dot \alpha} \Phi = 0
\ee
and has bosonic components
\be
\Phi | &=& z , 
\\
D^2 \Phi|  &=& G.
\ee
A free chiral superfield is described by the Lagrangian
\be
\label{chiral}
{\cal L} = \int d^4 \theta \ \bar \Phi \Phi , 
\ee
which leads to the superspace equations of motion
\be
\bar D^2 \bar \Phi =0.
\ee
On the other hand a complex linear superfield \cite{Gates:1980az} is defined by the condition
\be
\label{CL}
\bar D^2  \S =0 , 
\ee
with superspace Lagrangian
\be
\label{linear}
{\cal L} = - \int d^4 \theta \ \bar \S \S , 
\ee
for the free theory.
The bosonic components of the complex linear are
\be
\S | &=& A , 
\\
D^2 \S|  &=& F , 
\\
\bar D_{\dot \beta} D_\alpha \S | &=& P_{\alpha \dot \beta}.
\ee
Note that this multiplet can be consistently coupled to supersymmetric Yang-Mills \cite{Penati:1997pm,Penati:1997fz}.

The equivalence between the two can be understood by using the following Lagrangian
\be
\label{dual}
{\cal L}_{\text{dual}} = - \int d^4 \theta \ \bar \S \S + \int d^4 \theta \ \S \Phi + \int d^4 \theta \ \bar \S \bar \Phi
\ee
for a chiral superfield $\Phi$ but an unconstrained $\S$.
Then, integrating out $\Phi$ gives a complex linear condition (\ref{CL}) on $\S$ and 
the dual Lagrangian (\ref{dual}) becomes (\ref{linear}), 
but, integrating out $\S$ we find 
\be
\bar \S = \Phi
\ee
and after we plug-back into (\ref{dual}) we recover the Lagrangian (\ref{chiral}).
For this reason  the chiral multiplet is commonly used: 
it is classically equivalent to the linear super multiplet and it is simpler to deal with. 
The propagators of the chiral and the complex linear superfields are given by
\be
\bar \Phi \Phi  : & \ \ \ &  \frac{D^2 \bar D^2}{p^2 } \ \delta^4(\theta - \theta') , 
\\
\bar \S \S  : & \ \ \ & \ls 1+   \frac{\bar D^2 D^2}{p^2 } \rs \delta^4(\theta - \theta').
\ee

Due to the structure of the complex linear multiplet it is only possible that it acquires a mass in 
tandem with a chiral multiplet \cite{Deo:1985ix}. 
This is done by modifying the complex linear constraint to 
\be
\label{mass}
\bar D^2 \S = m \Phi. 
\ee
The way to understand the constraint is by dualizing the massive chiral superfield Lagrangian
\be
\label{massive}
{\cal L} = \int d^4 \theta \, \bar \Phi \Phi 
- \frac{m}{2} \int d^2 \theta \, \Phi^2 
- \frac{m}{2} \int d^2 \bar \theta \, \bar \Phi^2 .
\ee
For an unconstrained superfield $\S$ and a chiral $\Phi$ we have  
\be
\nn
{\cal L}_{\text{dual}} &=& - \int d^4 \theta \ \bar \S \S 
+ \int d^4 \theta \ \S \Phi + \int d^4 \theta \ \bar \S \bar \Phi
\\
\label{dualmassive}
&&- \frac{m}{2} \int d^2 \theta \, \Phi^2 
- \frac{m}{2} \int d^2 \bar \theta \, \bar \Phi^2 .
\ee
Then by the equations of motion for $\S$ we go to (\ref{massive}). 
But, the equations of motion for $\Phi$ give (\ref{mass})
$$
\bar D^2 \S = m \Phi.
$$
If on top of these we use the $\S$ equation of motion $\bar \S = \Phi$ 
we find 
\be
\bar D^2 \S = m \bar \S , 
\ee
which is the equation of motion for a massive chiral superfield;  
in other words this is a chiral superfield in the disguise of a complex linear \cite{Buchbinder:1998qv}.
Thus it is not possible to have a pure massive complex linear because it will 
always be in fact a relabeling of the massive chiral multiplet. 
It is then clear that the condition (\ref{mass}) indeed represents a massive complex linear, 
but, it is not possible to solve this condition in terms of $\S$ alone, 
as we did before, thus completely disposing of the chiral. 
To have a truly massive complex linear superfield, the chiral one has to stay in the picture as well. 
The consistent way to write the massive Lagrangian is
\be
\label{linear-chiral-mass}
{\cal L} =  - \int d^4 \theta \ \bar \S \S + \int d^4 \theta \ \bar \Phi \Phi , 
\ee
with the condition (\ref{mass}) for $\S$.
The mass terms are not manifest in (\ref{linear-chiral-mass}), 
but they are of course there in the component form \cite{Deo:1985ix}. 
Imposing the condition (\ref{mass}) via a Lagrange multiplier \cite{GonzalezRey:1997xp} 
also makes the mass terms appear in superspace 
and then one can find the standard propagators for the massive theory 
\be
\bar \Phi \Phi  : & \ \ \ &  \frac{D^2 \bar D^2}{p^2 + m^2} \delta^4(\theta - \theta') , 
\\
\bar \S \S  : & \ \ \ & \ls 1+   \frac{\bar D^2 D^2}{p^2 + m^2} \rs \delta^4(\theta - \theta') , 
\ee
and the extra propagators due to the mass terms 
\be
\bar \S \Phi  : & \ \ \ &  \frac{ m \bar D^2}{p^2 + m^2} \delta^4(\theta - \theta') , 
\\
\bar \Phi \S  : & \ \ \ &  \frac{ m  D^2}{p^2 + m^2} \delta^4(\theta - \theta').
\ee
Note that this is in fact an ${\cal N} = 2$ hypermultiplet \cite{GonzalezRey:1997qh,GonzalezRey:1997xp}.

Finally, it is interesting what happens to a supersymmetry breaking theory with a linear superpotential under
the duality. We start from
\be
\label{break}
{\cal L} = \int d^4 \theta \, \bar \Phi \Phi 
- f \int d^2 \theta \, \Phi 
- f \int d^2 \bar \theta \, \bar \Phi , 
\ee
which leads to the equations of motion for $\Phi$  
\be
\label{chiralbreak}
\bar D^2 \bar \Phi = f , 
\ee
and for the auxiliary field specifically
\be
\label{chiralbreakG}
G = f , 
\ee
signaling supersymmetry breaking.
On the other hand it is again possible to dualize this Lagrangian. 
For an unconstrained superfield $\S$ and a chiral $\Phi$ we have  
\be
\nn
{\cal L}_{\text{dual}} &=& - \int d^4 \theta \ \bar \S \S 
+ \int d^4 \theta \ \S \Phi + \int d^4 \theta \ \bar \S \bar \Phi
\\
\label{dualbreak}
&&- f \int d^2 \theta \, \Phi 
- f \int d^2 \bar \theta \, \bar \Phi .
\ee
Then by the equations of motion for $\S$ we go to (\ref{break}). 
But, the equations of motion for $\Phi$ give \cite{Kuzenko:2011ti}
\be
\label{brno}
\bar D^2 \S = f , 
\ee
and the duality Lagrangian (\ref{dualbreak}) becomes 
\be
\label{free2}
{\cal L}= - \int d^4 \theta \ \bar \S \S.
\ee
The Lagrangian (\ref{free2}) with the constraint (\ref{brno}) imply supersymmetry breaking. 
Nevertheless, if one turns to components, all auxiliary fields get a vanishing vacuum expectation value.  
This peculiarity is resolved by turning back to (\ref{dualbreak}).  
Combining the equations of motion for $\Phi$, which are (\ref{brno}),  
with the $\S$ equations of motion 
\be
\bar \S = \Phi , 
\ee
one can see  again that (\ref{brno}) is in fact a relabeling of the chiral superfield supersymmetry breaking (\ref{chiralbreak}).
Thus, the constraint (\ref{brno}) already contains the information of supersymmetry breaking originating by an underlying  chiral superfield 
whose auxiliary field $G$ indeed gets a {\it vev} from (\ref{chiralbreakG}).

Mechanisms to break supersymmetry by a complex linear have recently been found, 
where indeed, the auxiliary field of the complex linear superfield gets a  {\it vev}  
and signals supersymmetry breaking. 
Consider the Lagrangian \cite{Farakos:2013zsa}
\be
\label{hd1}
{\cal L} = - \int d^4 \theta \, \S \bar \S 
+ \frac{1}{8 f^2}  \int d^4 \theta \, 
D^{\alpha} \S  D_{\alpha}  \S \bar D^{\dot \alpha} \bar  \S 
 \bar D_{\dot \alpha} \bar  \S .
\ee
The scalar auxiliary sector of this theory reads
\be
{\cal L}_{F} = -F \bar F + \frac{1}{2 f^2} F^2 \bar F^2   , 
\ee
leading to the equations of motion
\be
\label{F3}
F ( F \bar F - f^2 ) =0.
\ee
The equation (\ref{F3}) on top of the supersymmetric solution
\be
F=0 , 
\ee
has a second solution  
\be
\label{s1}
F \bar F =  f^2 , 
\ee 
which signals supersymmetry breakdown. 
It has been shown that for models of this sort \cite{Farakos:2013zsa},  
the goldstino will be one of the previously auxiliary fermion fields of the theory, 
which propagates only in the broken branch.
Its supersymmetry transformation is
\be
\delta \lambda \sim \langle F \rangle \epsilon .
\ee
One may also assume the existence of superspace higher derivatives 
of higher dimension on top of (\ref{hd1}),  
for example 
\be
\label{F3}
{\cal L}_{F} = -F \bar F + \frac{1}{2 f^2} F^2 \bar F^2   +\frac{1}{ {f'}^4} F^3 \bar F^3  + \cdots
\ee 
and investigate if the solution (\ref{s1}) is still valid even though it has been found by ignoring the $1/f'$ term. 
There are two limiting cases where conclusions can be drawn. 
First consider 
\be
\label{no-hierarchy}
\frac{1}{ f} \lesssim \frac{1}{ f'} \ \  \leftrightarrow \ \  \frac{f}{f'} \gtrsim 1 . 
\ee
In this case (\ref{F3})  becomes 
\be
{\cal L}_{F} = - \frac12 f^2 + f^2 \left( \frac{f}{ f'} \right)^4   + \cdots 
\ee
which means that it is inconsistent to consider the solution  (\ref{s1}) 
since the terms we ignored are larger than the terms we took into account in order to find the solution. 
The other limiting case is the existence of a large hierarchy 
between the two scales 
\be
\label{hierarchy}
\frac{1}{ f} >> \frac{1}{ f'}   \ \  \leftrightarrow \ \  \frac{f}{f'} << 1 . 
\ee
Then (\ref{F3})  becomes 
\be
{\cal L}_{F} = - \frac12 f^2 + f^2 \left( \frac{f}{ f'} \right)^4   + \cdots 
\ee
thus the higher order terms are highly suppressed 
and may be safely ignored. 
In fact in a generic effective theory higher order terms as in (\ref{F3}) will also be generated, 
and the verification or not of  the hierarchy (\ref{hierarchy}) 
is a clear signal for the existence or not of the new vacua. 
Of course depending on the mechanism responsible for the generation of these superspace higher derivatives  
one will in principle find different results.

\subsection{Gauge multiplets and gauging}

A massive $U(1)$ vector multiplet is described by the Lagrangian 
\be
\label{massiveU1}
{\cal L}=  \int d^2 \theta\; W^{\alpha} W_{\alpha} + h.c. 
+ \frac{1}{2}  M^2 \int d^4\theta\, V^2,
\ee
with $V = \bar V$ and
\be
W_{\alpha} = i \bar D^2 D_{\alpha} V.
\ee
The bosonic components of this multiplet are
\be
V | &=& C,
\\
D^2 V | &=& N,
\\
\frac{1}{2} [\bar D_{\dot \alpha} , D_{\alpha}] V | &=& A_{\alpha \dot \alpha},
\\
\frac{1}{2} D^\alpha \bar D^2 D_\alpha  V | &=& \text{D}.
\ee
One may imagine that the massive $U(1)$ in (\ref{massiveU1}) has acquired a mass 
via a gauge invariant mechanism, for example in interaction with a real linear multiplet $L$, 
as described in Ref. \cite{CFG87,Siegel:1979ai}. 
In that case we have the Lagrangian
\be
\begin{split}
\label{massiveU2}
{\cal L}=& \int d^2 \theta W^{\alpha} W_{\alpha} + h.c. 
\\
&- \frac{1}{2}  \int d^4 \theta L^2
+ M  \int d^4 \theta L V,
\end{split}
\ee
which is invariant under
\be
\label{gauge}
V \rightarrow V + i \bar \Lambda - i \Lambda,
\ee
due to 
\be
 \int d^4 \theta L\, \Lambda = \int d^4 \theta L\, \bar \Lambda = 0 ,
\ee
for a chiral superfield 
$\Lambda$ and real linear $L$.
Note that by dimensional analysis $[V]=0$ and $[L]=1$, thus indeed $[M]=1$.
One can then rewrite (\ref{massiveU2}) as 
\be
\begin{split}
\label{massiveU3}
{\cal L}= & \int d^2 \theta W^{\alpha} W_{\alpha} + h.c. 
- \frac{1}{2}  \int d^4 \theta L^2
\\
&+ M  \int d^4 \theta L V 
+  \int d^4 \theta L (S + \bar S),
\end{split}
\ee
where now $L$ is real but otherwise unconstrained and $S$ is a chiral multiplet which renders $L$ linear on shell.
Since now $L$ is unconstrained, we can integrate it out to find
\be
\begin{split}
\label{massiveU4}
{\cal L}= &  \int d^2 \theta W^{\alpha} W_{\alpha} + h.c. 
\\
& + \frac{1}{2}  \int d^4\theta\, (M V + S + \bar S)^2,
\end{split}
\ee
which after a redefinition of $V$ of the form (\ref{gauge}) with $\Lambda = i S$ becomes (\ref{massiveU1}).
Here the vector superfield has {\it eaten} the real linear and became massive.

\begin{widetext}
Since in Lagrangian (\ref{massiveU1}) we have a massive $U(1)$, 
it is possible to invert the propagator without gauge-fixing. 
We will nevertheless introduce a gauge fixing term parametrized by $\xi$ 
\be
\label{GF}
{\cal L}_{GF}= - \xi \int d^4 \theta D^2 V \bar D^2 V ,
\ee
which gives the final Lagrangian
\be
\begin{split}
\label{massiveU1+GF}
{\cal L}=   \int d^2 \theta W^{\alpha} W_{\alpha} + h.c. 
- \xi \int d^4 \theta D^2 V \bar D^2 V 
+ \frac{1}{2}  M^2 \int d^4\theta\, V^2,
\end{split}
\ee
and the propagator of the massive vector superfield is
\be
V V  :  \ \ \  \left( - \frac{1}{p^2 + M^2}  \frac{D \bar D^2 D}{p^2} 
+  \frac{\xi^{-1}}{p^2 + \xi^{-1} M^2}  \frac{D^2 \bar D^2 +\bar D^2 D^2}{p^2}  \right)   \delta^4(\theta - \theta') .
\ee
One can now have three different forms for the propagator, depending on the value of $\xi$ 
\begin{itemize}

\item

For no gauge fixing ($\xi = 0$) the propagator becomes
\be \label{nG}
V V  :  \ \ \  \left( - \frac{1}{p^2 + M^2}  \frac{D \bar D^2 D}{p^2} 
+  \frac{1}{ M^2}  \frac{D^2 \bar D^2 +\bar D^2 D^2}{p^2}  \right)   \delta^4(\theta - \theta')   .
\ee

\item

In the Feynman gauge ($\xi = 1$) it becomes
\be \label{FG}
\hskip-5.5cm
V V  :  \ \ \   -\frac{1}{p^2 + M^2} \delta^4(\theta - \theta').
\ee

\item

In the Landau gauge ($\xi = \infty$) it becomes
\be \label{LG}
\hskip-4.2cm
V V  :  \ \ \   - \frac{1}{p^2 + M^2}  \frac{D \bar D^2 D}{p^2}  \delta^4(\theta - \theta').
\ee

\end{itemize}
\end{widetext}
We will perform our calculations in the Feynman gauge. 
Note that in the Landau gauge (\ref{LG}) and the Feynman gauge (\ref{FG}) it is consistent to set $M=0$, 
but for the propagator without gauge fixing (\ref{nG}) taking the limit $M \rightarrow 0$ 
results into a singularity, as expected.

The gauging of a chiral multiplet is well known to be
\be
\label{BCGC}
{\cal L} = \int d^4 \theta \, \bar \Phi e^V \Phi,
\ee
where  the vector transforms as shown in (\ref{gauge}) and the chiral superfield transforms as
\be
\Phi \rightarrow \Phi' = e^{i \Lambda} \Phi ,
\ee
for a chiral $\Lambda$. 
It is easy to verify that the transformed superfield is still chiral
\be
\bar D_{\dot \alpha} \Phi' = \bar D_{\dot \alpha}  ( e^{i \Lambda} \Phi) =0. 
\ee
The gauging of the complex linear is 
\be
\label{BSGS}
{\cal L} = - \int d^4 \theta \, \bar \S e^V \S,
\ee
but note that the gauge transformation of $\S$ is again 
\be
\S \rightarrow \S' = e^{i \Lambda} \S ,
\ee
where $\Lambda$ is still chiral.
It is easy to verify that the transformed superfield remains "complex linear"
\be
\bar D^2 \S' = \bar D^2 ( e^{i \Lambda} \S) =e^{i \Lambda}  \bar D^2 ( \S) =  0. 
\ee
On the other hand, a transformation of the form 
\be
\label{SG}
\S \rightarrow \S' = e^{i S} \S ,
\ee
for a complex linear $S$ would violate the complex linearity of the $\S$ superfield 
due to the fact that 
\be
\bar D^2 \, S^n \ne 0   \ , \ \text{for} \ \ n>1 .
\ee
A transformation (\ref{SG}) would also lead to a second complication. 
The invariance of (\ref{BSGS}) would require
\be
\label{gaugeS}
V \rightarrow V + i \bar S - i S .
\ee
But the field strength of the gauge superfield
\be
W_{\alpha} = i \bar D^2 D_{\alpha} V,
\ee
is not invariant under (\ref{gaugeS})  since
\be
i \bar D^2 D_{\alpha} (-i S + i \bar S ) \ne 0.
\ee

\section{Gauged massless complex linear and massive $U(1)$}

We start with a model of a massless complex linear coupled to a massive $U(1)$. 
We calculate the effective low energy theory for the scalar multiplet 
by integrating out the massive vector superfield.
We find that the effective theory on top of the corrections to the 
K\"ahler potential also contains superspace higher derivatives which lead to 
the deformations of the auxiliary potential.

\begin{figure}[tbp] 
\includegraphics[width = 0.2 \textwidth]{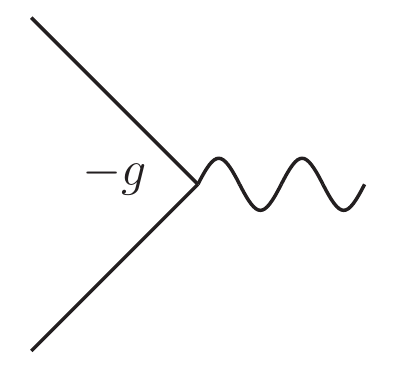}
\caption{Vertex due to gauging.}
\label{1A}
\end{figure}

\begin{figure}[tbp]
\includegraphics[width = 0.2 \textwidth]{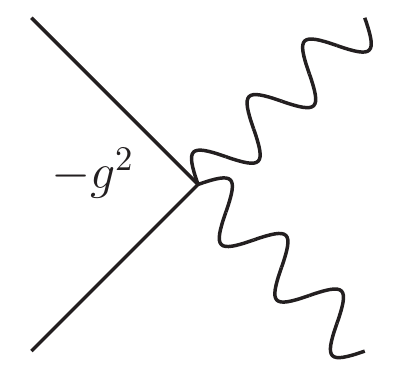}
\caption{Vertex due to gauging.}
\label{1B}
\end{figure}

Our theory has the following form in superspace 
\be
\begin{split}
\label{L1}
{\cal L}=& -\int d^4\theta\, \bar{\S} e^{g V} \S 
+  \int d^2 \theta W^{\alpha} W_{\alpha} + h.c. 
\\
&- \xi \int d^4 \theta D^2 V \bar D^2 V 
+ \frac{1}{2}  M^2 \int d^4\theta\, V^2,
\end{split}
\ee
which is easily shown to be classically equivalent to a gauged chiral superfield theory 
coupled to a massive vector superfield by gauging the Lagrangian (\ref{dual}) 
for the dual superfields having opposite $U(1)$ charge.
Here the gauge transformation for $\S$ is
\be
\S \rightarrow e^{i g \Lambda} \S.
\ee
The model (\ref{L1}) gives rise to the vertices shown in Fig. (\ref{1A}) and  Fig. (\ref{1B})  (up to ${\cal O}(g^2)$). 
\begin{figure}[tbp] 
\includegraphics[scale = 0.8]{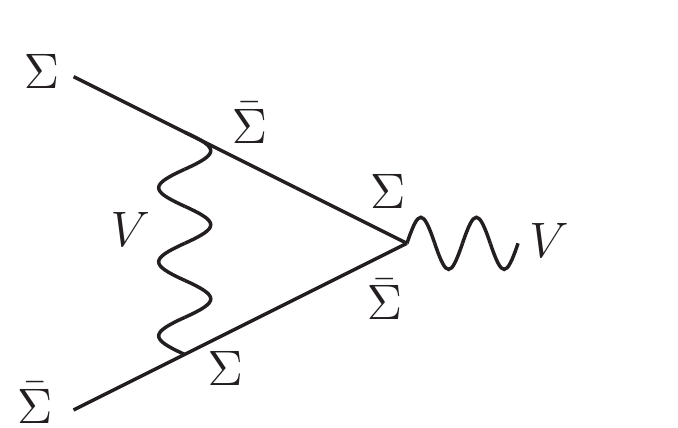}
\caption{One-loop correction to the $\S\, \bar\S\, V$ vertex.}
\label{2A}
\end{figure}
\begin{figure}[tbp] 
\includegraphics[scale = 0.75]{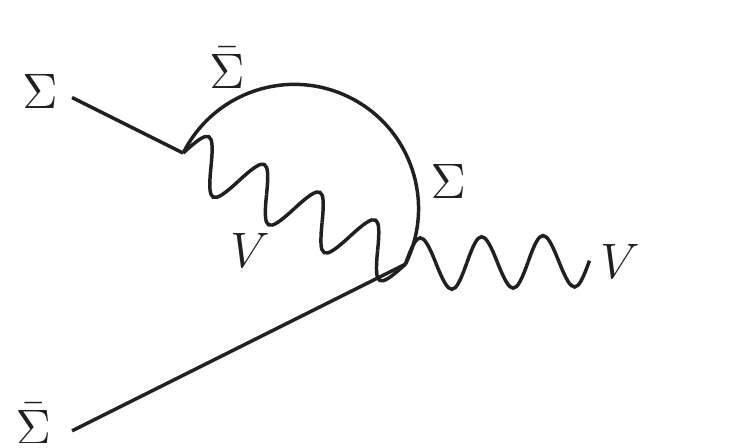}
\caption{One-loop correction to the $\S\, \bar\S\, V$ vertex.}
\label{2B}
\end{figure}
\noindent
The renormalizability is understood in terms of gauge invariance, 
and the no tree-level supersymmetry breaking by a straightforward component expansion. 
Indeed, the scalar non-derivative sector of the Lagrangian (\ref{L1}) is
\be
\label{VAC}
{\cal L} = - F \bar F + N \bar N + \text{D} ( -g A \bar A + M^2 C )+ (1-  \xi ) \text{D}^2 \!,
\ee
which has no other vacuum than
\be
\langle F\rangle = \langle N\rangle = \langle \text{D}\rangle = 0.
\ee

\noindent
To calculate the vacuum structure of the tree-level theory (Eq.(\ref{VAC})) 
we have written the gauge invariant terms in the Wess-Zumino gauge.  
Of course when calculating the radiative corrections one should 
{\it not} turn to the W-Z gauge for the quantum fluctuations 
even though it is possible to do so for the background vector superfield (see for example Ref. \cite{Grisaru:1982df}).

To find the low energy effective theory for $\S$  we work in two steps
\begin{enumerate}

\item We write down the 1-loop corrections to the vertices.

\item Then we integrate out the massive vector superfield.

\end{enumerate}
Here we will keep terms up to 
\be
{\cal O}(\frac{1}{M^4}),
\ee
since this is the scale where the auxiliary field potential deformation  comes in. 
Due to the massless complex linear multiplet, 
the diagrams under consideration will have infrared divergences, 
thus we will calculate the loop momenta integrals with an IR cut-off
\be
\Lambda : \text{IR cut-off}.
\ee
Moreover, since for the moment we are only interested 
in finding contributions to the scalar potential, 
we  will set the external momenta to zero, 
or equivalently, impose the condition  
\be
\p_a \S =  \p_a \bar \S = 0 ,
\ee
which is allowed by supersymmetry.

\begin{figure}[tbp] 
\includegraphics[scale = 0.75]{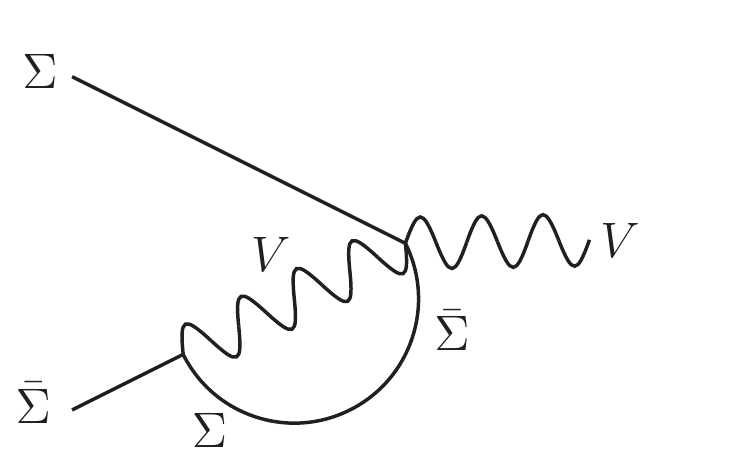}
\caption{One-loop correction to the $\S\, \bar\S\, V$ vertex.}
\label{2C}
\end{figure}

\begin{figure}[tbp] 
\includegraphics[width= 0.3 \textwidth]{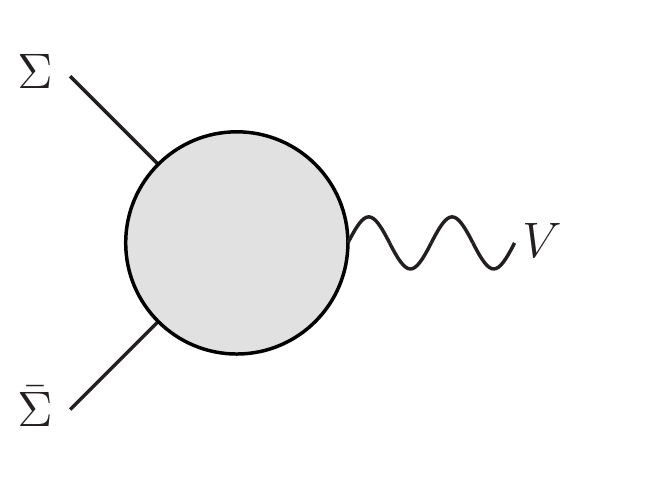}
\caption{Tree-level and one-loop corrections to the $\S\, \bar\S\, V$ vertex.}
\label{3}
\end{figure}

\begin{figure}[tbp] 
\includegraphics[width= 0.35 \textwidth]{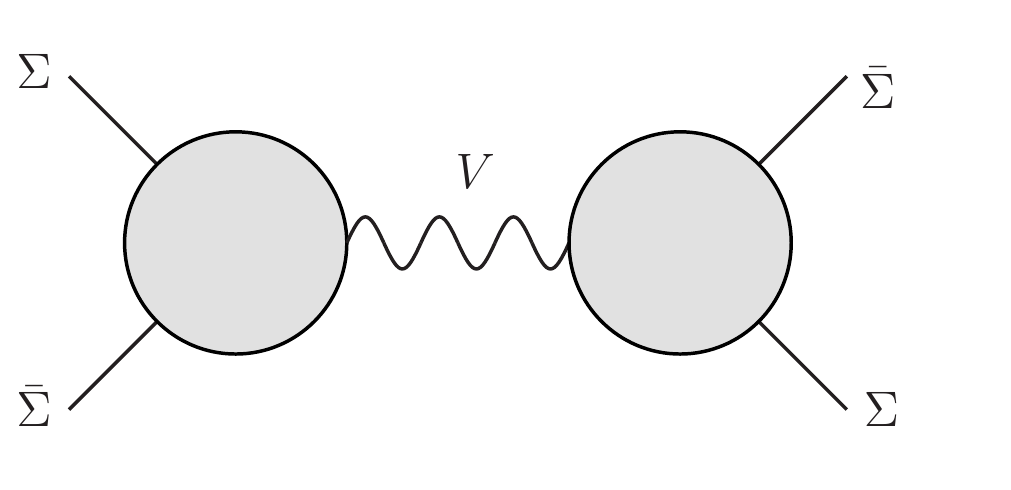}
\caption{Effective interactions from integrating out $V$.}
\label{4A}
\end{figure}

\begin{figure}[tbp] 
\includegraphics[width= 0.35 \textwidth]{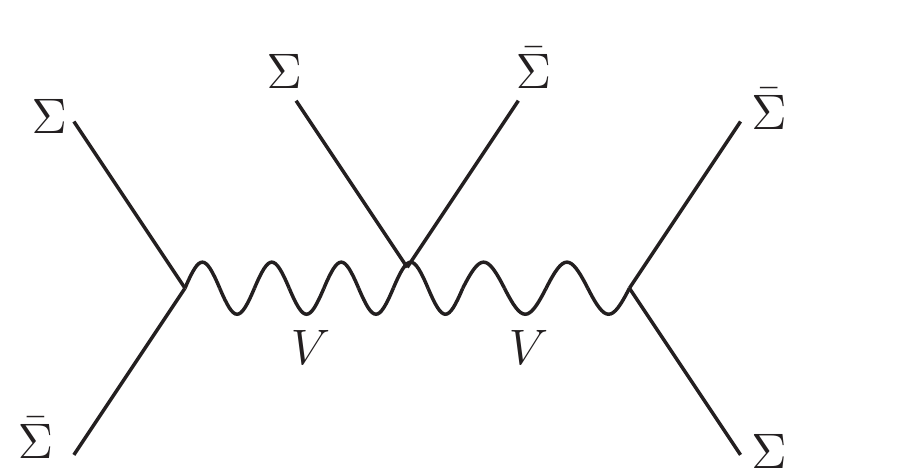}
\caption{Effective interactions from integrating out $V$.}
\label{4B}
\end{figure}

\begin{figure}[tbp] 
\includegraphics[scale = 0.75]{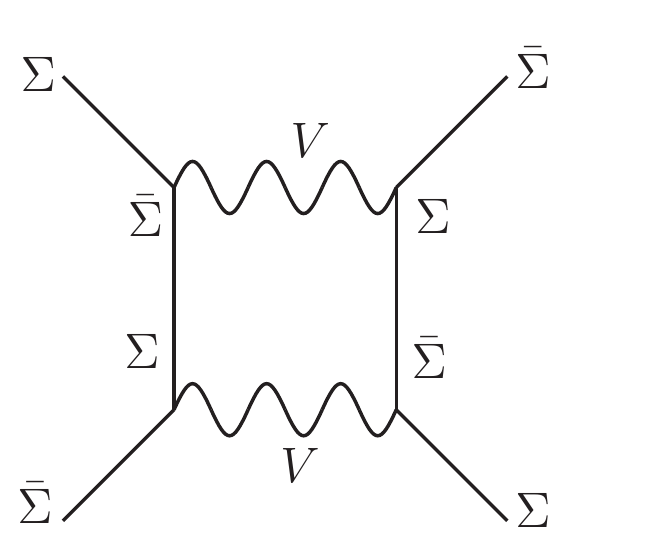}
\caption{One-loop correction to the $\S^2\, \bar\S^2$ vertex.}
\label{5A}
\end{figure}

\begin{figure}[tbp] 
\includegraphics[scale = 0.75]{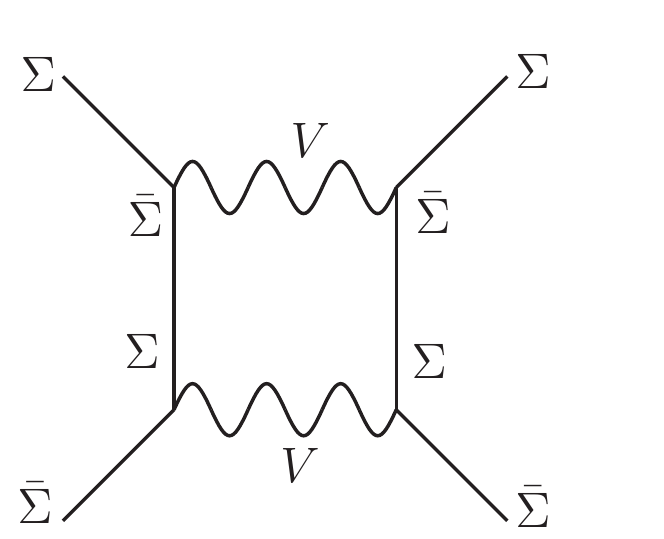}
\caption{One-loop correction to the $\S^2\, \bar\S^2$ vertex.}
\label{5B}
\end{figure}

\begin{figure}[tbp] 
\includegraphics[scale = 0.75]{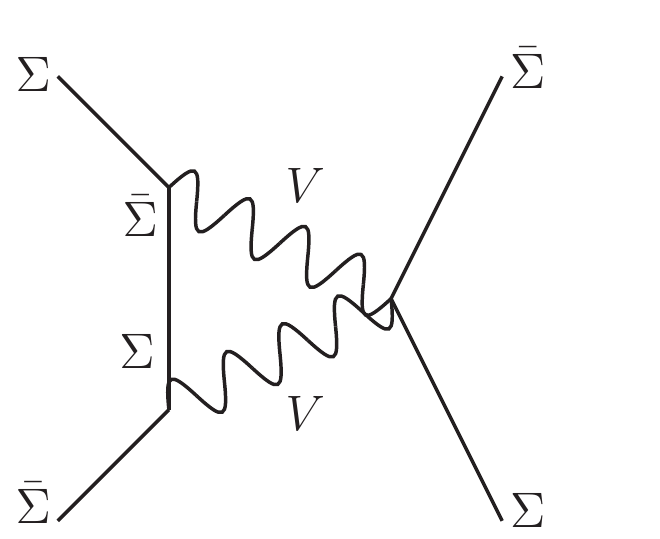}
\caption{The 1-loop correction to the $\S^2 \, \bar \S^2$ vertex.}
\label{5C}
\end{figure}

The one-loop corrections to the $V \S \bar \S$ vertex are given by the diagrams in Fig. (\ref{2A}),  (\ref{2B}) 
and  (\ref{2C})
\noindent
with corresponding terms in the effective theory
\be
\begin{split}
{\cal L}_{3} =&  \frac{ g^3}{16 \pi^2 } \,  {\cal A}(\epsilon) \int d^4 \theta \, \S \bar \S V 
\\
& + \frac{g^3}{16 \pi^2 M^2}  \text{ln}\left( \frac{\Lambda^2 + M^2}{\Lambda^2 }\right)  \int d^4 \theta  \S \bar \S  D^2 \bar D^2 V  ,
\\
{\cal L}_{4} =& - \frac{ g^3}{16 \pi^2 }  \,  {\cal A}(\epsilon) \int d^4 \theta \, \S \bar \S V ,
\\
{\cal L}_{5} =&  -\frac{ g^3}{16 \pi^2 }  \,  {\cal A}(\epsilon) \int d^4 \theta \, \S \bar \S V .
\end{split}
\ee
The factor $\,  {\cal A}(\epsilon)$ contains the poles due to the UV divergencies 
and finite terms
\be
{\cal A}(\epsilon) = \frac{2}{\epsilon} +1 -\gamma + \text{ln} (4 \pi)  - \text{ln}\left( \frac{\Lambda^2 + M^2}{ \mu^2} \right)  .
\ee
We have introduced the renormalization scale $\mu$ in order for $g$ to remain dimensionless 
\be
g \rightarrow \mu^{\epsilon/2} g ,
\ee
during the regularization. 
Then we have the diagram shown in Fig. (\ref{3}) which represents 
a vertex containing the tree level vertex (\ref{1A}) along with the loop corrections (\ref{2A}-\ref{2C}),  
\noindent
and can be written in the effective theory as 
\be
\begin{split}
{\cal L}_{3}  = & \int d^4 \theta ( - g \bar \S \, V \, \S ) + {\cal L}_{2(a)} + {\cal L}_{2(b)} + {\cal L}_{2(c)} 
\\
= & \int d^4 \theta \Big{(} - g    \bar \S \, V \, \S  
\\
&   + \frac{g^3}{16 \pi^2 M^2}  \text{ln}\left( \frac{\Lambda^2 
+ M^2}{\Lambda^2 }\right)  \S \bar \S  D^2 \bar D^2 V    \Big{)} ,
\end{split}
\ee
where we have absorbed the finite parts of the loop corrections to the tree-level vertex  by redefining $g$. 
Since we want to integrate out the vector superfield,  
the leading diagrams to this order due to the vertex corrections, 
relevant to our results, are shown in the Fig. (\ref{4A}) and  (\ref{4B})  
\noindent
and lead to the effective interactions 
\be
\begin{split}
{\cal L}_{7} =& -\frac{g^2}{2 M^2} \int d^4 \theta \, \S^2 \bar \S^2  
\\
 & + \frac{g^4}{16 \pi^2 M^4}  \text{ln}\left( \frac{\Lambda^2 + M^2}{\Lambda^2 } \right)  \int d^4 \theta \, D^2 (\S \bar \S) \bar D^2 (\S \bar \S )  
\end{split}
\ee
and
\be
{\cal L}_{8} = -\frac{g^4}{6} \int d^4 \theta \frac{1}{M^4} \S^3 \bar \S^3 .
\ee
The one-loop corrections to the four sigma vertex terms are given by the diagrams in Fig. (\ref{5A}),  (\ref{5B}) and  (\ref{5C}), 
\noindent
with corresponding terms
\be
\begin{split}
{\cal L}_{9} =&  \frac{g^4}{32 \pi^2 (M^2 + \Lambda^2)} \int d^4 \theta \, \S^2 \bar \S^2
\\
& +  \frac{g^4}{32 \pi^2 M^4}   \int d^4 \theta   
 \lc  \text{ln}\left(\frac{\Lambda^2 + M^2}{\Lambda^2 }\right) \right.
\\
&\left.-\frac{M^2}{(\Lambda^2 + M^2)} \rc 
 D^2 ( \S \bar \S )   \bar D^2 ( \S \bar \S ) ,
\\
{\cal L}_{10} =&  \frac{g^4}{16 \pi^2 (M^2 + \Lambda^2)} \int d^4 \theta \, \S^2 \bar \S^2
\\
& +  \frac{g^4}{32 \pi^2 M^4}   \int d^4 \theta   
 \lc  \text{ln}\left(\frac{\Lambda^2 + M^2}{\Lambda^2 }\right)  \right.
\\
& \left.-\frac{M^2}{(\Lambda^2 + M^2)} \rc 
 D^2 (  \bar \S^2 )   \bar D^2 ( \S^2 ) 
\end{split}
\ee
and
\be
{\cal L}_{11} &=& - \frac{g^4}{16 \pi^2 (M^2 + \Lambda^2)} \int d^4 \theta \, \S^2 \bar \S^2 .
\ee

\begin{figure}[tbp] 
\includegraphics[width = 0.35 \textwidth]{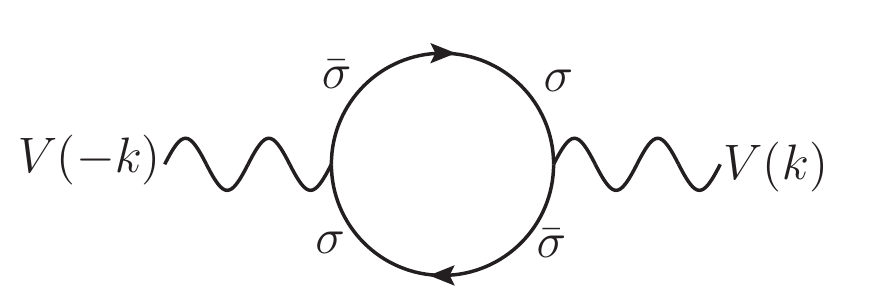}
\caption{One-loop correction to the vector propagator from the $\s$ superfield.}
\label{6A}
\end{figure}

Gathering  the quantum corrections up to order $1 / M^4$
we have  for $\S$
\be
\begin{split}
{\cal L}_{\S,eff} =& - {\cal T} \int d^4\theta\, \bar{\S}  \S
 - {\cal P} \int d^4 \theta  \S^2 \bar \S^2 
- {\cal Q}  \int d^4 \theta  \S^3 \bar \S^3
\\
\label{Seff}
& +  {\cal R} \int d^4 \theta D^2 ( \S  \bar \S )   \bar D^2 ( \S  \bar \S ) 
\\
& +  {\cal S} \int d^4 \theta D^2 ( \bar \S^2 )   \bar D^2 ( \S^2 ).
\end{split}
\ee
with
\be
{\cal T} &=& 1 + {\cal O}(g^2),
\\
{\cal P} &=& \frac{g^2}{2 M^2} + {\cal O}(g^4),
\\
{\cal Q} &=& \frac{g^4}{6 M^4}  + {\cal O}(g^6) , 
\ee
and
\be
\begin{split}
{\cal R} =&  \frac{ g^4}{16 \pi^2 M^4} \lc 
 2\,  \text{ln}\left(\frac{\Lambda^2 + M^2}{\Lambda^2 }\right) 
-\frac{M^2}{(\Lambda^2 + M^2)} \rc + {\cal O}(g^6),
\\
{\cal S} =&  \frac{g^4}{32 \pi^2 M^4} \lc   \text{ln}\left(\frac{\Lambda^2 + M^2}{\Lambda^2}\right) 
-\frac{M^2}{(\Lambda^2 + M^2)} \rc + {\cal O}(g^6).
\end{split}
\ee
We see that  superspace higher derivatives of similar form as those  introduced in Ref. \cite{Farakos:2013zsa} 
are indeed generated by quantum corrections.   
The scalar sector of the  Lagrangian (\ref{Seff}) is
\be
\begin{split}
{\cal L}_{\S,eff}^{scalar} = &  
- F \bar F \ls  {\cal T} + 4 \, {\cal P} \, A \bar A + 9 \, {\cal Q} \, A^2 \bar A^2 \rs
\\
&+  F^2 \bar F^2 \, {\cal R} ,
\end{split}
\ee 
where the impact of the superspace higher derivatives on the auxiliary field's effective potential is manifest.

\section{Gauged massive hypermultiplet and massive $U(1)$}

A massive complex linear $\S$ is defined by the supersymmetric condition 
\be
\bar{D}^2 \S= m \Phi,
\ee
where $\Phi$ is a chiral superfield. 
Thus $\S$ acquires a mass $m$ in tandem with the chiral multiplet $\Phi$ \cite{Deo:1985ix}. 
It is in fact equivalent to a massive $N=2$ hypermultiplet \cite{GonzalezRey:1997xp}.
In a gauged $U(1)$ model, 
both $\S$ and $\Phi$ transform with the same charge
\be
\S \rightarrow e^{i g \Lambda} \S \ , \ \ \Phi \rightarrow e^{i g \Lambda} \Phi.
\ee
\begin{figure}[tbp] 
\includegraphics[width = 0.35  \textwidth]{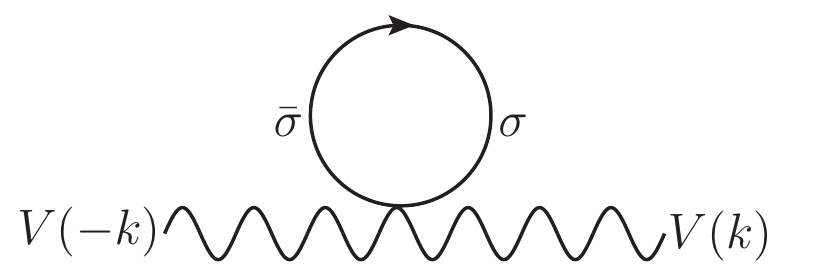}
\caption{One-loop correction to the vector propagator from the $\s$ superfield.}
\label{6B}
\end{figure}
\begin{figure}[tbp] 
\includegraphics[width =  0.35 \textwidth]{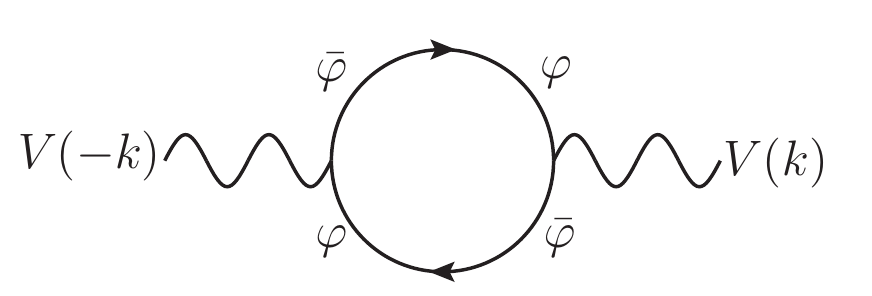}
\caption{One-loop correction to the vector propagator from the $\varphi$ superfield.}
\label{7A}
\end{figure}
\noindent
We employ the following superspace Lagrangian
\be
\begin{split}
{\cal L}= & -\int d^4\theta\, \bar{\S} e^{g V} \S 
+\int d^4\theta\, \bar{\Phi} e^{g V} \Phi
\\
&
+ \int d^2 \theta W^{\alpha} W_{\alpha}(V) + h.c. 
\\
\label{maas}
& \  - \xi  \int d^4 \theta D^2 V \bar D^2 V 
+ \frac{1}{2} M^2  \int  d^4\theta\, V^2.
\end{split}
\ee
The scalar non-derivative component sector of theory (\ref{maas}) is
\be
\begin{split}
{\cal L} =& - F \bar F + N \bar N + \text{D} (g z \bar z -g A \bar A + M^2 C ) 
\\
&+ (1- \xi )  \text{D}^2 -m^2 z \bar z - m A \bar G - m \bar A G + G \bar G,
\end{split}
\ee
with a supersymmetric vacuum
\be
\langle F\rangle = \langle G\rangle = \langle N\rangle = \langle\text{D}\rangle =  0.
\ee

We now turn to the quantum corrections. 
A way to rewrite the complex linear multiplet constraint is by splitting $\S$ into
a background field $\S_0$ and the quantum fluctuations  $\s$ as follows
\be
\S = \S_0 +\s.
\ee
For the gauge multiplet we consider no background field 
\be
V = 0 + V,
\ee
and for the chiral superfield we have 
\be
\Phi = 0 + \varphi,
\ee 
where
\be
\bar{D}^2 \s= M \varphi 
\ee
and
\be
\bar{D}^2 \S_0= 0.
\ee
Here we have set the hypermultiplet mass equal to the vector multiplet mass for simplicity and 
we will keep it like this. 
This is not merely technical, since if these masses originate from the same physics, they will also be of the same order.

\subsection{Propagator renormalization}

\begin{figure}[tbp] 
\includegraphics[width =  0.35 \textwidth]{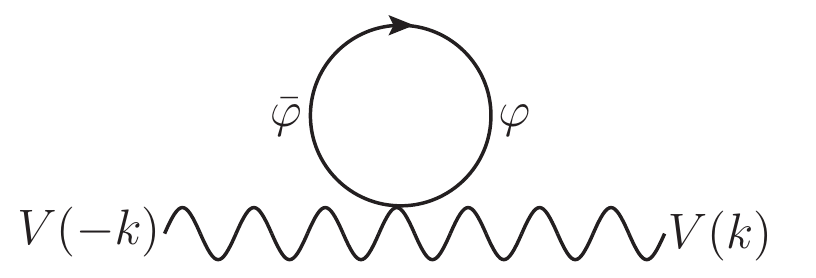}
\caption{One-loop correction to the vector propagator from the $\varphi$ superfield.}
\label{7B}
\end{figure}

The corrections to the vector propagator due to $\s$ loops are given by the diagrams in Fig. (\ref{6A}) and (\ref{6B}) 
and contribute to the effective theory as 
\be
\begin{split}
{\cal L}_{12} =& \frac{g^2}{32\pi^2}\int d^4\theta \int d^4 k\;
V(-k)\left( -M^2-\frac{1}{6}k^2 \right.
\\
&\left. \ \ \ \ \ \ \ \ \ \ \ \ \ \ \ \ \  \ \ \ \ \ 
+\frac12 {\cal B}(\epsilon) D^\alpha \bar{D}^2 D_\alpha
\right) V(k),
\\
{\cal L}_{13} =&   \frac{g^2}{32\pi^2}\int d^4\theta \int d^4 k\;
V(-k)M^2({\cal B}(\epsilon)+1)V(k),
\end{split}
\ee
where
\be
{\cal B}(\epsilon) =  \frac{2}{\epsilon}  -\gamma + \text{ln} (4 \pi)  - \text{ln}\left( \frac{ M^2}{ \mu^2} \right)  .
\ee
Corrections  due to $\varphi$ loops are given by the diagrams shown in Fig. (\ref{7A}) and (\ref{7B}) 
\begin{figure}[tbp] 
\includegraphics[width = 0.33 \textwidth]{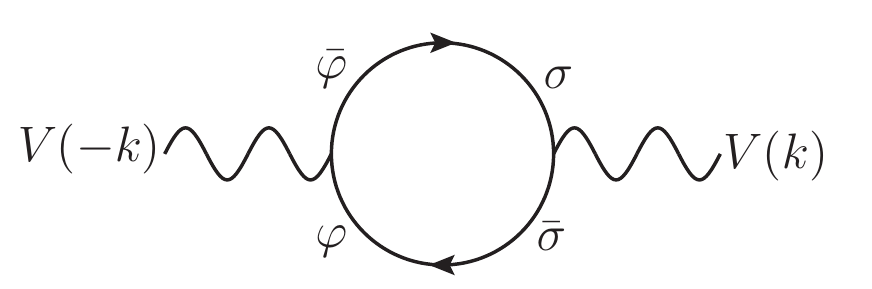}
\caption{One-loop correction to the vector propagator from the $\varphi$ and $\s$ superfields.}
\label{8}
\end{figure}
\noindent
and contribute to the effective theory as
\be
\begin{split}
{\cal L}_{14} \! &= \!
 \frac{g^2}{32\pi^2} \!\int \!\! d^4\theta \!\! \int \!\! d^4 k \,
V(-k)
\Big{(}
2{\cal B}(\epsilon)M^2 \!+\! M^2 \! -\frac{1}{6}k^2 \!
\\
& \ \ \ \ \ \  \ \ \ \ \ \   \ \ \ \ \ \   \ \ \ \ \ \ +\frac12 {\cal B}(\epsilon) D^\alpha \bar{D}^2 D_\alpha
V(k)
\Big{)},
\\
{\cal L}_{15} &=  -\frac{g^2}{32\pi^2}\int d^4\theta \int d^4 k\;
V(-k)M^2\left({\cal B}(\epsilon)+1\right)V(k) .
\end{split}
\ee
\begin{figure}[tbp] 
\includegraphics[width = 0.43 \textwidth]{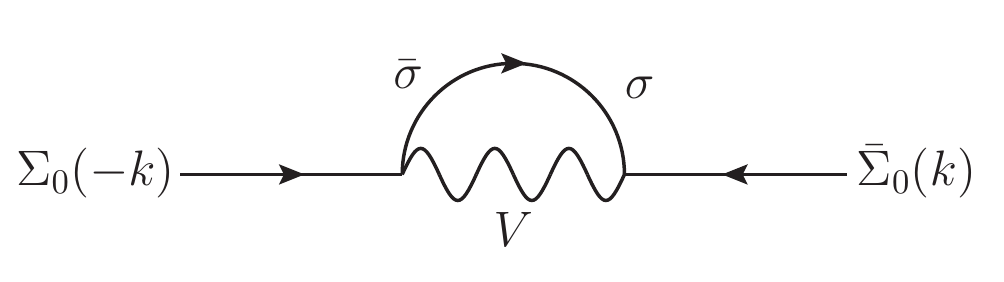}
\caption{One-loop correction to the background superfield.}
\label{9}
\end{figure}
\begin{figure}[tbp] 
\includegraphics[scale = 0.8]{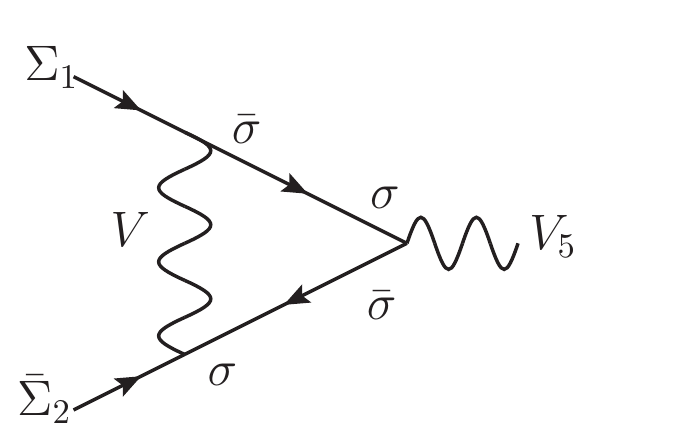}
\caption{One-loop correction to the $\S_0\, \bar\S_0\, V$ vertex.}
\label{10A}
\end{figure}

\noindent
There is  finally a mixed loop diagram shown in Fig. (\ref{8}), 
contributing to the effective theory as
\be
{\cal L}_{16} \!\!=\!\! - \frac{g^2}{32\pi^2} \!\!\int \!\! d^4\theta \!\! \int \!\! d^4 k \; \!
V\!(-k)\!\!\left(\!2{\cal B}(\epsilon)M^2 \!-\! \frac13 k^2\! \right)\! \! V\!(k) \!.
\ee
Summing up the different contributions one gets
\be
\begin{split}
{\cal L} &= {\cal L}_{12}+{\cal L}_{13}+{\cal L}_{14}+{\cal L}_{15} + {\cal L}_{16}
\\
&= \frac{g^2}{32\pi^2}\int d^4\theta \int d^4 k\;
V(-k) {\cal B}(\epsilon) D^\alpha \bar{D}^2 D_\alpha V(k) .
\end{split}
\ee
It is gratifying to see that the quadratic divergencies cancel 
and that the contribution is purely transversal as it should be for gauge invariance.

For the background complex linear we have the diagram in Fig. (\ref{9}) 
\be
{\cal L}_{17} \!\!=\!\!  -\frac{g^2}{16\pi^2}\!\!\int\!\! d^4\theta\!\! \int d^4 k\; \!
\S\!(-k)\!\! \left(\!{\cal B}(\epsilon)M^2\! -\! \frac16 k^2\!\right)\!\!\bar \S\!(k)\! 
\ee
which will give us the $\S_0$ wave function renormalization.

\subsection{Vertices and integrating out}

We now wish to integrate out all the massive fluctuations and find the low energy 
effective theory for the background field $\S_0$.
Again we  consider the gauge coupling $g$ to be small.
Note also that now the complex linear superfield is massive, there will be no IR divergence.

\begin{figure}[tbp] 
\includegraphics[scale = 0.7]{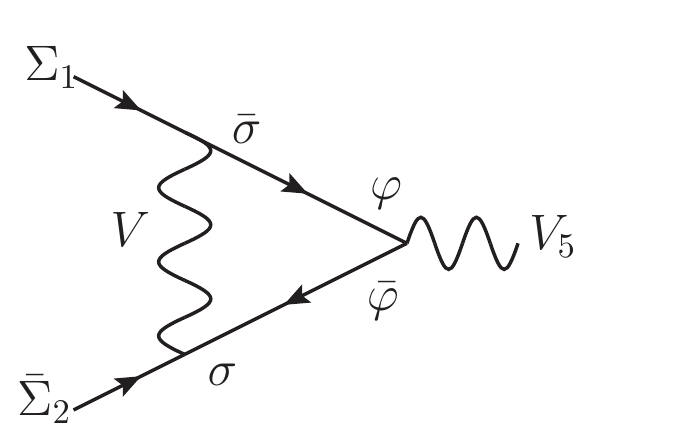}
\caption{One-loop correction to the $\S_0\, \bar\S_0\, V$ vertex.}
\label{10B}
\end{figure}

\begin{figure}[tbp] 
\includegraphics[width= 0.25 \textwidth]{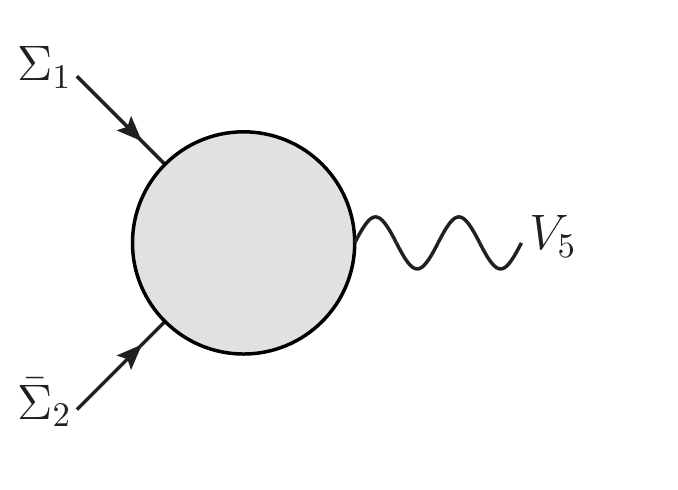}
\caption{Tree-level and one-loop corrections to the $\S_0 \bar \S_0 V$ vertex.}
\label{12}
\end{figure}

\begin{figure}[tbp] 
\includegraphics[width=0.35 \textwidth]{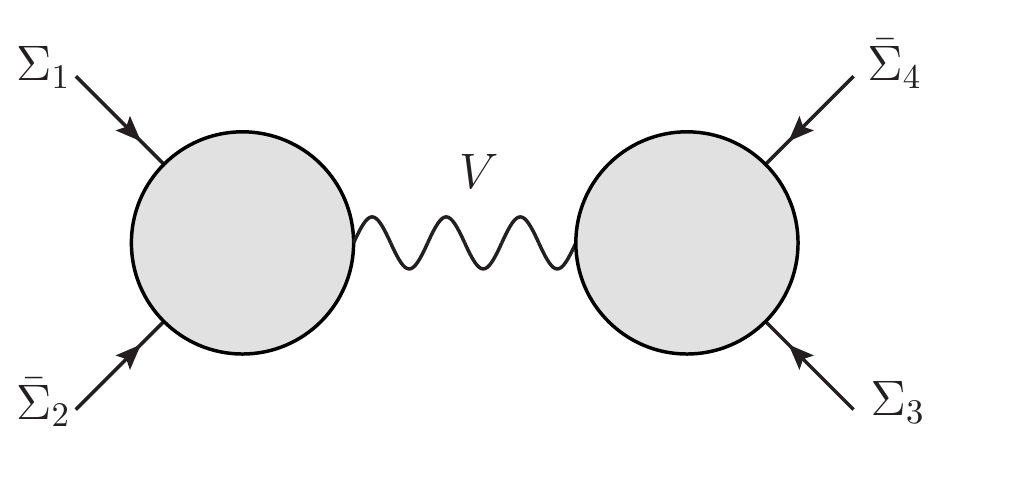}
\caption{Effective interactions due to integrating out $V$.}
\label{13A}
\end{figure}

\begin{figure}[tbp] 
\includegraphics[width= 0.3 \textwidth]{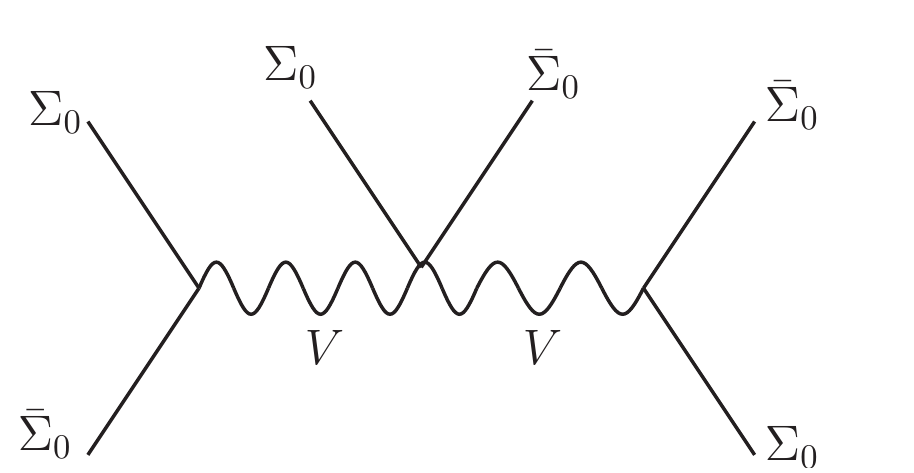}
\caption{Effective interactions due to integrating out $V$.}
\label{13B}
\end{figure}

\begin{figure}[t] 
\includegraphics[scale = 0.65]{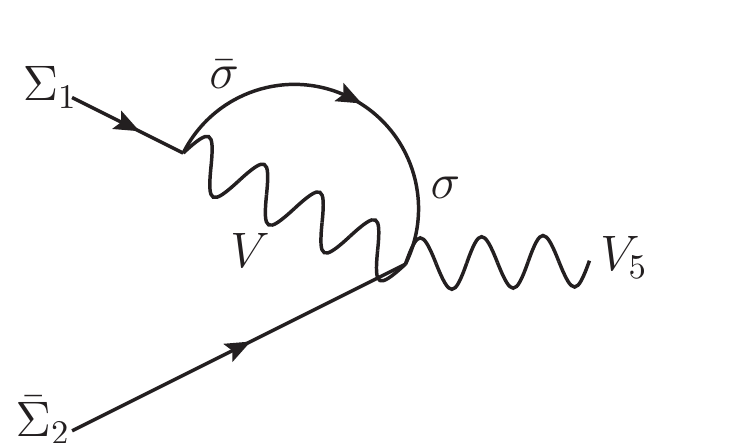}
\caption{One-loop correction to the $\S_0\, \bar\S_0\, V$ vertex.}
\label{11A}
\end{figure}

\begin{figure}[t] 
\includegraphics[scale = 0.7]{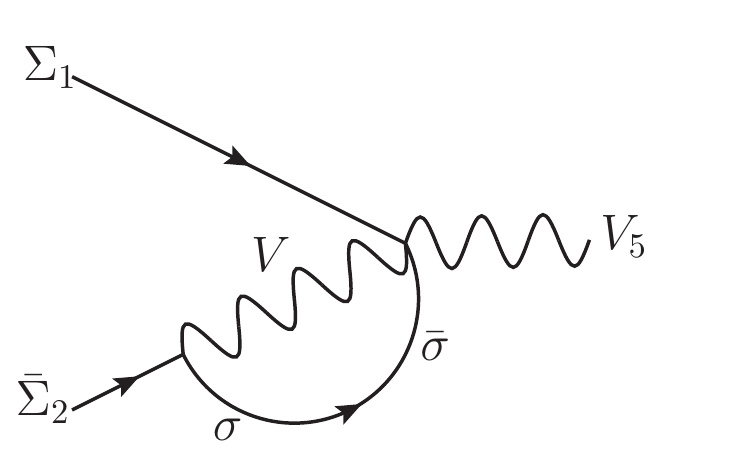}
\caption{One-loop correction to the $\S_0\, \bar\S_0\, V$ vertex.}
\label{11B}
\end{figure}

\begin{figure}[tbp] 
\includegraphics[scale = 0.7]{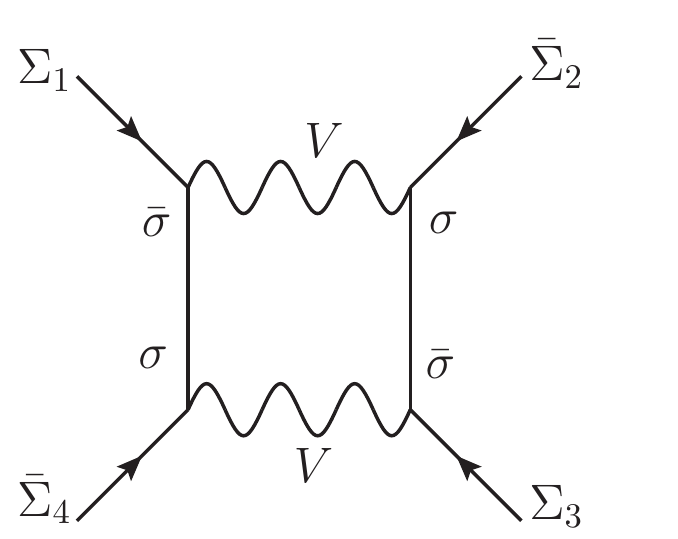}
\caption{One-loop correction to the $\S_0^2 \, \bar \S_0^2$ vertex.}
\label{14A}
\end{figure}

\begin{figure}[tbp] 
\includegraphics[scale = 0.7]{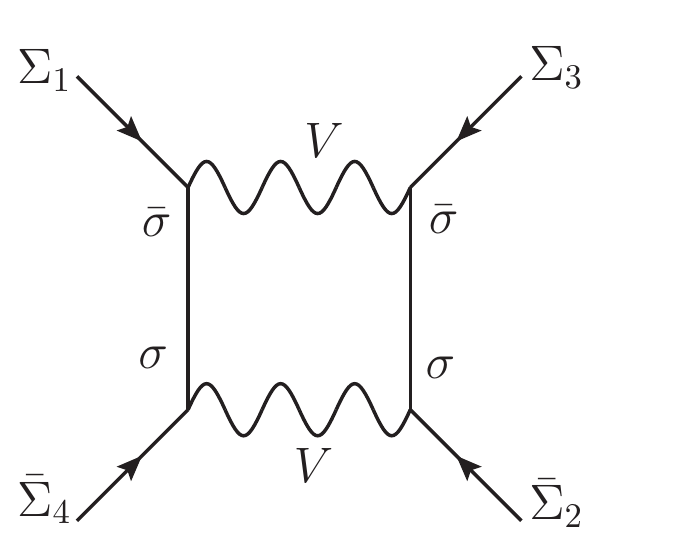}
\caption{One-loop correction to the $\S_0^2 \, \bar \S_0^2$ vertex.}
\label{14B}
\end{figure}

\begin{figure}[tbp] 
\includegraphics[scale = 0.7]{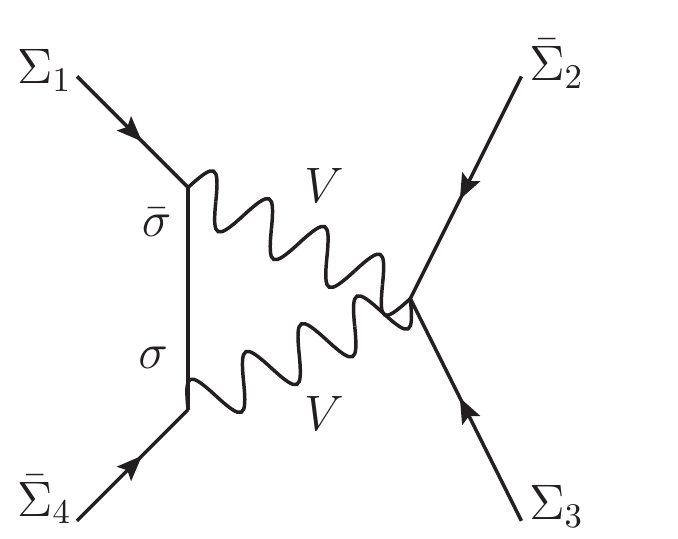}
\caption{One-loop correction to the $\S_0^2 \, \bar \S_0^2$ vertex.}
\label{14C}
\end{figure}

\begin{widetext}
The one-loop corrections to the $V \S \bar \S$ vertex are given by the diagrams in Fig. (\ref{10A}) and (\ref{10B}) 
with corresponding terms in the effective theory 
\be
\nn
{\cal L}_{18} &=& \! \frac{g^3}{32 \pi^2 M^2} \int \! \! d^4 \theta \! \! \int \! \! d^4 p_{\text{ext}} 
\, \S_1 \bar \S_2 \! 
\left( M^2 +2 M^2  {\cal B}(\epsilon) -\frac{1}{3} p_1^2-\frac{1}{6} p_1 p_2-\frac{1}{3} p_2^2  
  \right) V_5 
\\
\nn
&& + \frac{g^3}{32 \pi^2 M^2} \int \! \! d^4 \theta \! \! \int \! \! d^4 p_{\text{ext}}  \, \S_1 \bar \S_2 \! 
\left( - (\frac{1}{6} p_1 + \frac{1}{3} p_5)^{\alpha \dot \alpha} \bar D_{\dot \alpha} D_{\alpha} V_5  
- (\frac{1}{6} p_2 + \frac{1}{3} p_5)^{\alpha \dot \alpha} D_{\alpha} \bar D_{\dot \alpha} V_5  \right)
\\
&&
+ \frac{g^3}{32 \pi^2 M^2} \int \! \! d^4 \theta \! \! \int \! \! d^4 p_{\text{ext}} \left( \frac{1}{2} \S_1 \bar \S_2 
 \lc D^2 , \bar D^2 \rc  V_5 \right) , 
\\
{\cal L}_{19} &=& - \frac{g^3}{32 \pi^2 M^2} \int d^4 \theta \int \! \! d^4 p_{\text{ext}}  \, \S_1 \bar \S_2 V_5 
\left( M^2 -\frac{1}{6} p_1^2-\frac{1}{6} p_1 p_2-\frac{1}{6} p_2^2   \right) ,
\ee
where
\be
\int \! \! d^4 p_{\text{ext}} = (2 \pi)^4 \delta(\sum_{i} \, p_i) \prod_{i} \int \frac{d^4 p_i}{(2 \pi)^4} .
\ee

Here $p_i$ are the external momenta and the subscript $i$ on the superfields implies their momentum, 
for example
\be
\S_i = \int d^4 x \, \S_0 (x,\theta,\bar \theta) e^{- i x p_i} .
\ee 
In the above diagrams (and the subsequent) we have calculated the external momenta contribution 
in the limit 
\be
\frac{p_{\text{ext}}^2}{M^2} << 1.
\ee

We also have the diagrams shown in Fig. (\ref{11A}) and  (\ref{11B}) 
which lead to the terms 
\be
{\cal L}_{20} &=& \frac{g^3}{16 \pi^2 M^2} \int d^4 \theta \int \! \! d^4 p_{\text{ext}}  \, \S_1 \bar \S_2  V_5  
\left( -M^2 {\cal B}(\epsilon) + \frac{1}{6} p_1^2 \right) ,
\\
{\cal L}_{21} &=& \frac{g^3}{16 \pi^2 M^2} \int d^4 \theta \int \! \! d^4 p_{\text{ext}}  \, \S_1 \bar \S_2  V_5  
\left( -M^2 {\cal B}(\epsilon) + \frac{1}{6} p_2^2 \right) .
\ee

Then we have a diagram shown in Fig. (\ref{12}) which represents a vertex containing 
the tree level vertex (\ref{1A}) along with the loop corrections (\ref{10A} - \ref{11B}) ,   
and can be written in the effective theory as 
\be
\nn
{\cal L}_{22}  &=& \int d^4 \theta ( - g \bar \S_0 \, V \, \S_0 ) + {\cal L}_{18} + {\cal L}_{19} + {\cal L}_{20} 
+ {\cal L}_{21} 
\\ 
\nn
&=& \int d^4 \theta  ( - g \bar \S_0 V \S_0 )  
+ \frac{g^3}{32 \pi^2 M^2} \int d^4 \theta \! \! \int \! \! d^4 p_{\text{ext}} 
\, \S_1 \bar \S_2 
\left( \frac{1}{6} p_1^2 + \frac{1}{6} p_2^2  
+ \frac{1}{2} \lc D^2 , \bar D^2 \rc  \right) V_5 
\\
\nn
&& \! + \frac{g^3}{32 \pi^2 M^2} \int \! \! d^4 \theta \! \! \int \! \! d^4 p_{\text{ext}}  \, \S_1 \bar \S_2 \! \! 
\left( - (\frac{1}{6} p_1 + \frac{1}{3} p_5)^{\alpha \dot \alpha} \bar D_{\dot \alpha} D_{\alpha} V_5  
- (\frac{1}{6} p_2 + \frac{1}{3} p_5)^{\alpha \dot \alpha} D_{\alpha} \bar D_{\dot \alpha} V_5  \right) 
\\ 
\nn
&=& \int d^4 \theta  ( - g \bar \S_0 V \S_0 )  
+ \frac{g^3}{32 \pi^2 M^2} \int d^4 \theta \! \! \int \! \! d^4 p_{\text{ext}} 
\, \S_1 \bar \S_2 
\left( \frac{1}{6} p_1^2 + \frac{1}{6} p_2^2  
+ \frac{1}{2} \bar D  D^2 \bar D  \right) V_5 
\\
&& \! + \frac{g^3}{32 \pi^2 M^2} \int \! \! d^4 \theta \! \! \int \! \! d^4 p_{\text{ext}}  \, \S_1 \bar \S_2 \! \! 
\left( - \frac{1}{12} ( p_1 - p_2)^{\alpha \dot \alpha} [ D_{\alpha} , \bar D_{\dot \alpha} ] V_5  \right) ,
\ee
where we have absorbed the finite parts of the loop corrections   by redefining $g$. 
Since we want to integrate out the vector superfield,  
the leading diagrams to this order due to the vertex corrections, 
relevant to our results, are shown in  Fig. (\ref{13A}) and (\ref{13B}) 
and lead to the effective interactions 
\be
\nn
{\cal L}_{23} &=& -\frac{g^2}{2  M^2} \int d^4 \theta \, \S_0^2 \bar \S_0^2  
+ \frac{g^4}{64 \pi^2 M^4} \int d^4 \theta \! \! \int \! \! d^4 p_{\text{ext}} 
\, \S_1 \bar \S_2 \bar D  D^2 \bar D  (\S_3 \bar \S_4) 
\\
\nn
&&
+ \frac{g^2}{4 M^4} \int d^4 \theta \! \! \int \! \! d^4 p_{\text{ext}} 
\, \S_1 \bar \S_2   \S_3 \bar \S_4 \lc (p_1 + p_2)^2 + (p_3 + p_4)^2   \rc
\\
\nn
&&
+ \frac{g^4}{32 \pi^2 M^4} \int d^4 \theta \! \! \int \! \! d^4 p_{\text{ext}} 
\, \S_1 \bar \S_2 
\left( \frac{1}{12} p_1^2 + \frac{1}{12} p_2^2  + \frac{1}{12} p_3^2 + \frac{1}{12} p_4^2  \right) (\S_3 \bar \S_4) 
\\
\nn
&& \! + \frac{g^4}{64 \pi^2 M^4} \int \! \! d^4 \theta \! \! \int \! \! d^4 p_{\text{ext}}  \, \S_1 \bar \S_2 \! \! 
\left( - \frac{1}{12} ( p_1 - p_2)^{\alpha \dot \alpha} [ D_{\alpha} , \bar D_{\dot \alpha} ] (\S_3 \bar \S_4)   \right) 
\\
&& \! + \frac{g^4}{64 \pi^2 M^4} \int \! \! d^4 \theta \! \! \int \! \! d^4 p_{\text{ext}}  \, \S_3 \bar \S_4 \! \! 
\left( - \frac{1}{12} ( p_3 - p_4)^{\alpha \dot \alpha} [ D_{\alpha} , \bar D_{\dot \alpha} ] (\S_1 \bar \S_2)   \right) 
\ee
and
\be
{\cal L}_{24} = -\frac{g^4}{6 M^4} \int d^4 \theta   \S_0^3 \bar \S_0^3 . 
\ee 
The one-loop corrections to the $\S_0^2 \bar \S_0^2$ vertex are given by the diagrams in 
Fig. (\ref{14A}),  (\ref{14B}),  (\ref{14C}),  
for which we have
\be
\nn
{\cal L}_{25} &=&  \frac{g^4}{32 \pi^2 M^4}   \int \! \! d^4 \theta \! \int \! \! d^4 p_{\text{ext}}  
 \,  \S_1  \bar \S_2    
\left( \frac{2}{3}  M^2 - \frac{1}{48} \left(p_1^2+p_2^2+p_3^2+p_4^2 \right) \right.
\\
&& \ \ \ \ \ \ \ \ \ \ \ \ \ \ \ \ \ \ \ \ \ \ \ \ \ \ \ \ \ \ \ \ \ \ \ \ \ \ \   \left. 
- \frac{23}{120}(p_1+p_2)^2-\frac{1}{60}(p_1+p_3)^2
\right.
\\
\nn
&& \ \ \ \ \ \ \ \ \ \ \ \ \ \ \ \ \ \ \ \ \ \ \ \ \ \ \ \ \ \ \ \ \ \ \ \ \ \ \   \left. 
+ \frac{1}{6} \bar D^2 D^2 - \frac{1}{48} \left( p_2+p_4   
\right)^{\alpha \dot \alpha} \left[D_{ \alpha}, \bar D_{\dot\alpha}\right] \right) \S_3  \bar \S_4  ,
\\
\nn
{\cal L}_{26} &=&  \frac{g^4}{32 \pi^2 M^4}   \int \! \! d^4 \theta \!\! \int \! \! d^4 p_{\text{ext}}  
  \, \! \S_1  \S_3    \!
\left(  M^2 - \frac{1}{12} p_1^2- \frac{1}{12} p_1 p_2- \frac{1}{12} p_2^2 \right.
\\
&& \ \ \ \ \ \ \ \ \ \ \ \ \ \ \ \ \ \ \ \ \ \ \ \ \ \ \ \ \ \ \ \ \ \ \ \ \ \ \   \left. 
\!\!\! - \frac{1}{12} p_3^2- \frac{1}{12} p_3 p_4- \frac{1}{12} p_4^2 
+ \frac{1}{6} \bar D^2 D^2 \right) \bar \S_2  \bar \S_4  , 
\\
{\cal L}_{27} &=& -  \frac{g^4}{32 \pi^2 M^4}   \int d^4 \theta \int \! \! d^4 p_{\text{ext}}  
    \,  \S_1 \bar \S_2 \,  \S_3 \bar \S_4    
\left(  M^2 - \frac{1}{6} p_1^2- \frac{1}{6} p_1 p_4- \frac{1}{6} p_4^2 \right) ,
\ee
leading to the effective interactions
\be
\begin{split}
 {\cal L}_{25}+{\cal L}_{26}+{\cal L}_{27} = & 
 \frac{g^4}{32 \pi^2 M^4}   \int \! \! d^4 \theta \! \int \! \! d^4 p_{\text{ext}}
  \,  \S_1  \bar \S_2
\left( \frac{2}{3}  M^2 - \frac{1}{48} \left(p_1^2+p_2^2+p_3^2+p_4^2 \right) \right.
\\
& \ \ \ \ \ \ \ \ \ \ \ \ \ \ \ \ \ \ \ \ \ \ \ \ \ \ \ \ \ \ \ \ \ \ \ \ \ \ \   \left. 
- \frac{23}{120}(p_1+p_2)^2-\frac{1}{60}(p_1+p_3)^2
\right.
\\
& \ \ \ \ \ \ \ \ \ \ \ \ \ \ \ \ \ \ \ \ \ \ \ \ \ \ \ \ \ \ \ \ \ \ \ \ \ \ \   \left. 
+ \frac{1}{3} \bar D^2 D^2 - \frac{1}{48} \left( p_2+p_4   
\right)^{\alpha \dot \alpha} \left[D_{ \alpha}, \bar D_{\dot\alpha}\right] \right) \S_3  \bar \S_4  . 
\end{split}
\ee
\end{widetext}


Now we can calculate the leading contributions to the off-shell effective potential of the theory. 
As we see it receives contributions both from the effective K\"ahler potential 
but also from the deformation of the auxiliary field potential. 
Gathering the relevant quantum corrections 
we have  for the background superfield $\S_0$
\be
\nn
{\cal L}_{\S_0,eff} &=& -  {\cal T'} \int d^4\theta\, \bar{\S}_0  \S_0
 - {\cal P'} \int d^4 \theta  \S_0^2 \bar \S_0^2 
\\
\nn
&&- {\cal Q'}  \int d^4 \theta  \S_0^3 \bar \S_0^3
\\
\label{S0}
&& 
+  {\cal R'} \int d^4 \theta D^2 ( \S_0  \bar \S_0 )   \bar D^2 ( \S_0  \bar \S_0 )
\\
\nn
&&+  {\cal S'} \int d^4 \theta D^2 ( \bar \S_0^2 )   \bar D^2 ( \S_0^2) , 
\ee
with
\be
 {\cal T'} &=& 1 + {\cal O}(g^2),
\\
{\cal P'} &=& \frac{g^2}{2 M^2} + {\cal O}(g^4),
\\
{\cal Q'} &=& \frac{g^4}{6 M^4}  + {\cal O}(g^6),
\ee
and for the superspace higher derivative operators
\be
\label{R'}
{\cal R'} &=&  \frac{7 g^4}{192 \pi^2 M^4}   + {\cal O}(g^6),
\\
{\cal S'} &=&  \frac{g^4}{192 \pi^2 M^4}  + {\cal O}(g^6).
\ee 
Note that in Lagrangian (\ref{S0}) various superspace higher derivatives have been   generated radiatively.


\subsection{Vacuum structure}

Let us focus on the scalar non-derivative sector of  Lagrangian (\ref{S0}) which reads
\be
\begin{split}
{\cal L}_{\S_0,eff}^{scalar} = &
- F_0 \bar F_0 \ls  {\cal T'} + 4 \, {\cal P'} \, A_0 \bar A_0 + 9 \, {\cal Q'} \, A_0^2 \bar A_0^2 \rs  
\\
&+  F_0^2 \bar F_0^2 \, {\cal R'}  .  
\end{split}
\ee 
We see that the effective theory has an intriguing similarity to the models of Ref. \cite{Farakos:2013zsa}. 
Indeed, the equation of motion for $F_0$ has two solutions
\begin{itemize}

\item Standard branch: 
\be
F_0=0 . 
\ee

\item Broken branch: 
\be
 F_0 \bar F_0 =  \frac{1}{2 {\cal R'}}  \ls   {\cal T'} 
+ 4 \, {\cal P'} \, A_0 \bar A_0 + 9 \, {\cal Q'} \, A_0^2 \bar A_0^2  \rs . 
\ee

\end{itemize}

The scalar potential of the broken branch will have the form
\be
\label{V}
{\cal V} = \frac{1}{4 {\cal R'}}  \ls   {\cal T'} + 4 \, {\cal P'} \, A_0 \bar A_0
 + 9 \, {\cal Q'} \, A_0^2 \bar A_0^2  \rs^2 , 
\ee
leading to a positive vacuum energy, and a supersymmetry breaking scale
\be
\label{FbarF}
\langle F_0 \bar F_0\rangle = \frac{{\cal T'}}{2 {\cal R'}} = \frac{96 \pi^2 M^4}{7 g^4} .
\ee

\begin{figure}[tbp] 
\includegraphics[width= 0.35 \textwidth]{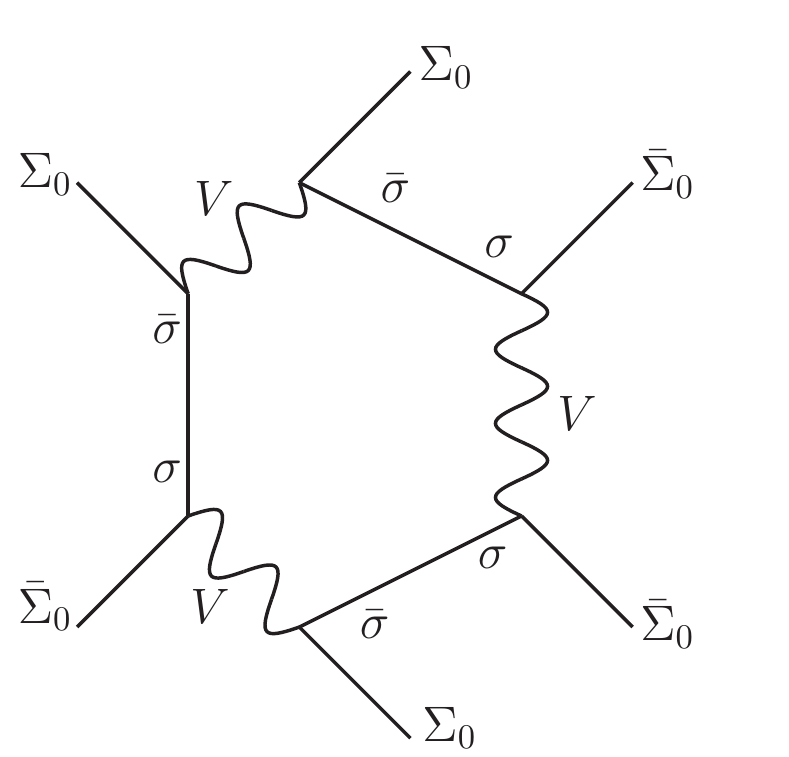}
\caption{Higher order diagram.}
\label{15A}
\end{figure}

\begin{figure}[tbp] 
\includegraphics[width= 0.35 \textwidth]{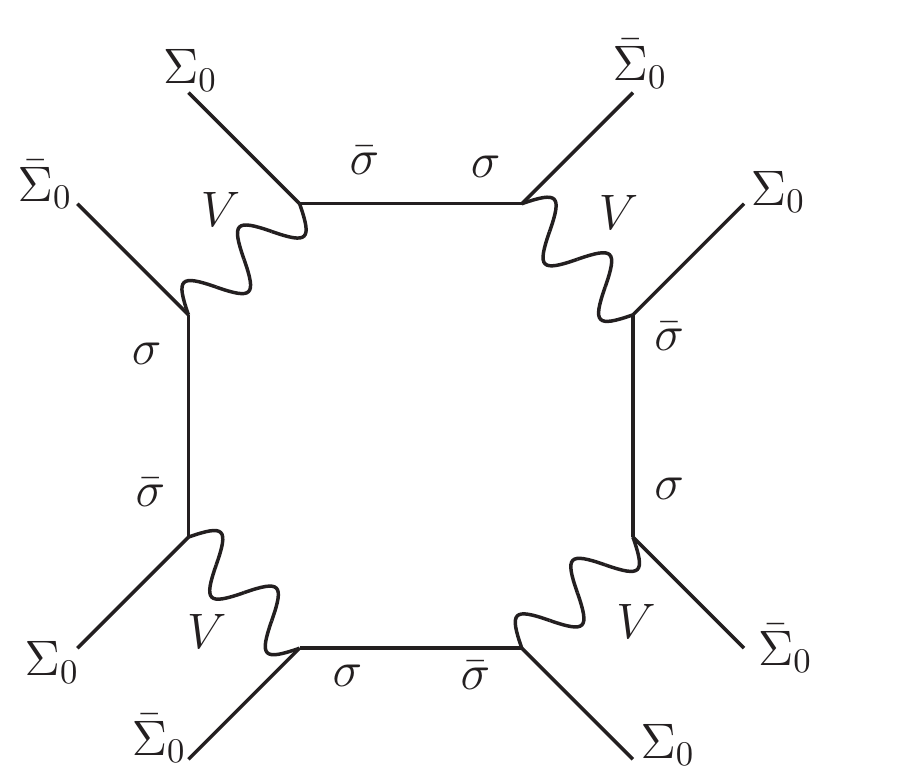}
\caption{Higher order diagram.}
\label{15B}
\end{figure}

Note that in this new vacuum $|F_0|$ has a dependence on $M^2/g^2$ which gives rise to the  question of higher order corrections. 
In other words, this solution can only be trusted if there exists a hierarchy between 
the leading superspace higher derivatives and the subsequent ones.  
If one naively estimates the one-loop contribution to higher point functions such as the 
six and eight point graphs of Fig. (\ref{15A}) and (\ref{15B}) one sees that the minimum of the potential is shifted. Namely,
the contribution from the 2n-point diagram to the effective potential of the auxiliary field goes as
\be\label{npcont}
(-1)^n M^4 \left(\frac{g^2F\bar F}{M^{4}}\right)^n . 
\ee
The auxiliary field potential will have a generic form 
\be
\label{higher-order-LF}
{\cal L}_F =M^4 h \left( \frac{g^2 F \bar F}{M^4}\right) - F \bar F,
\ee
but from (\ref{npcont}) we see that the higher order terms  do not satisfy 
the hierarchy criterion (\ref{hierarchy}) required for reliable results on supersymmetry breaking, 
and a complete knowledge of the form of (\ref{higher-order-LF})  would in principle be required.


Even though we do not have an  exact form for the higher order corrections, 
we can still 
ask if  generically supersymmetry is broken in the low energy limit 
by studying the renormalization group flow. 
We can draw  reliable results if we study a generic term of the from 
\be
\int d^4 \theta \lambda^{(j)} (D \S)^2  (\bar D \bar \S)^2 (D^2 \S \,  \bar D^2 \bar \S)^j , 
\ee 
which could arise from the quantum corrections, 
where the dimension of $\lambda^{(j)}$ is $-4(j+1)$.

Let us define the bare action as 
\be
\begin{split}
{\cal L}_{\rm bare} = \int d^4 \theta & \left(  V_b D^\alpha \bar D^2 D_\alpha V_b + M_b^2 V_b^2 \right.
\\
& \left. - \S_b \bar \S_b - g_b V_b \S_b \bar \S_b \right) 
\\
= \int d^4 \theta & \left( {\cal Z}_V  V D^\alpha \bar D^2 D_\alpha V +  M^2 V^2 \right.
\\
& \left. - {\cal Z}_\S  \S \bar \S - {\cal Z}_\alpha g V \S \bar \S \right) , 
\end{split}
\ee
from where we find 
\be
\begin{split}
{\cal Z}_\S &= 1 - \frac{4 \alpha}{\epsilon} , 
\\
{\cal Z}_V &= 1 - \frac{2 \alpha}{\epsilon} , 
\\
{\cal Z}_\alpha &= 1 - \frac{4 \alpha}{\epsilon} , 
\end{split}
\ee
with 
\be
\alpha = \frac{g^2}{32 \pi^2} . 
\ee
Since the bare coupling
\be
\alpha_b = \mu^\epsilon \alpha \frac{{\cal Z}_\alpha^2 }{{\cal Z}_\S^2 {\cal Z}_V} , 
\ee
 is independent of the scale $\mu$ we have 
\be
\label{ra} 
\frac{d \text{ln} \alpha}{d \text{ln} \mu}  =  - \epsilon + 2 \alpha  + {\cal O }(\alpha^2) , 
\ee
giving $\beta_\alpha =2 \alpha$ . 
Similarly we find 
\be
\label{dM}
\frac{1}{M} \frac{d M}{d \text{ln} \mu} =  \alpha  + {\cal O }(\alpha^2)  . 
\ee


The bare action for the higher dimension operators is defined as
\be
\int d^4 \theta \lambda_b^{(j)} (D \S_b)^2  (\bar D \bar \S_b)^2 (D^2 \S_b \,  \bar D^2 \bar \S_b)^j . 
\ee 
The wave function renormalization gives us a relation between the bare and the renormalized $\lambda$ 
\be
\lambda_b^{(j)} = \lambda^{(j)} ({\cal Z}_\S)^{-j -2} . 
\ee
In terms of the  dimensionless parameter $\tilde \lambda^{(j)} = \lambda^{(j)} \mu^{(j+1)(4-\epsilon)}$ 
we have 
\be
\lambda_0^{(j)} = \tilde \lambda^{(j)} ({\cal Z}_\S)^{-j -2} \mu^{(j+1)(\epsilon - 4)} , 
\ee
which gives
\be
\frac{d {\rm ln} \tilde \lambda^{(j)} }{d {\rm ln} \mu} = 4 (j+1) + 4 (j + 2 ) \alpha . 
\ee

We now want to study the emergence of the hierarchy. 
This translates into comparing $\left( \lambda^{(0)} \right)^{\frac14}$ to $\left( \lambda^{(j)} \right)^{\frac{1}{4(j+1)}}$ 
in the low energy limit. 
We find
\be
\label{run}
\frac{d}{d {\rm ln} \mu} \lc  
\frac14  {\rm ln} \left( \lambda^{(0)} \right)
\! - \! 
\frac{1}{4(j+1)}  {\rm ln}  \left( \lambda^{(j)} \right) 
 \rc \! = \!  \frac{j}{j+1} \alpha . 
\ee 
From formula (\ref{run})  
we conclude that even though the quantum corrections generate the superspace higher derivatives responsible 
for supersymmetry breaking, 
they do not generate the required hierarchy between the leading and the subsequent terms  
and thus the 
solution leading to the 
broken branch can not be trusted.  
In other words, 
the quantum corrections alone can not lead to the supersymmetry breaking. 
Our results show that in the case where supersymmetry is broken 
by the mechanism of  Ref. \cite{Farakos:2013zsa}, 
these superspace higher derivatives have to be related to the 
underlying theory or  rely on some other mechanism to be generated.

\noindent

Note  
that in a different setup where $\beta_\alpha < 0$ in formula (\ref{ra}), 
the model would have the opposite behavior in the low energy, 
leading to a hierarchy and a reliable supersymmetry breaking branch.

\section{Discussion}

In this work we have studied low energy effective theories for complex linear superfields. 
We have calculated the quantum corrections to the effective action 
including also the superspace higher derivative terms on top of the usual corrections to 
the K\"ahler potential. 
This was done by calculating  tree-level and one-loop quantum corrections, 
and then integrating out the massive sector.

Our motivation was  related to the properties of such operators concerning supersymmetry breaking.  
We underlined that a hierarchy between the higher dimension operators is essential 
for the supersymmetry breaking vacua to be consistent. 
Turning to the effective theory we have verified that indeed these operators are generated by the radiative corrections. 
On the other hand the required hierarchy between the leading terms and the subsequent ones sufficient 
for supersymmetry breaking was not found. 
This leads us to conclude that if supersymmetry is broken by the specific superspace higher derivatives, 
these terms have to originate from the underlying theory or another mechanism with different 
IR properties for the beta function.

We close with a comment on the case where one does not integrate out the massive modes. 
There the auxiliary field deformation terms  for all the multiplets have to be taken  into account. 
The theory, after the radiative corrections are introduced, 
would be of the form ({\it e.g.} for a complex linear and a vector superfield)
\be
\label{concl}
{\cal L} \supset   \text{D}^2 - F \bar F + \frac{1}{M^4} \text{D}^4  +\frac{1}{M^4} F^2 \bar F^2 \cdots  
\ee
The study of the vacuum structure of a theory like (\ref{concl})  is left  for future work.

\section*{Acknowledgements}

We  thank I. Bakas, N. Irges, A. Kehagias and U. Lindstr\"om for discussion and correspondence. 
This work was supported by the Grant agency of the Czech republic under the grant P201/12/G028.

\end{document}